\documentclass[preprint,showpacs,showkeys,amsmath,amssymb,12pt,a4paper]{revtex4-1}
\usepackage[utf8]{inputenc}
\usepackage{graphicx}
\usepackage{amsmath}
\usepackage{amssymb}
\usepackage{color}
\usepackage{hyperref}

\begin{document}

\title{Lepton Number Violation, Lepton Flavour Violation and Baryogenesis in Left-Right Symmetric Model}
\author{Happy Borgohain}
\email{happy@tezu.ernet.in}
\author{Mrinal Kumar Das}
\email{mkdas@tezu.ernet.in}
\affiliation{Department of Physics, Tezpur University, Tezpur 784028, India}

\begin{abstract}

 We did a model independent phenomenological study of baryogenesis via leptogenesis, neutrinoless double beta decay (NDBD) and charged lepton flavour violation (CLFV)
 in a generic left-right symmetric model 
 (LRSM) where neutrino mass originates from the type I + type II seesaw mechanism. We  studied the new physics contributions to NDBD coming from the left-right gauge boson mixing and the heavy
 neutrino contribution within the framework of LRSM. We have considered the mass of the RH  gauge boson to be specifically 5 TeV, 10 TeV and 18 TeV
and studied the effects of the new physics contributions on the effective mass and baryogenesis and compared with the current experimental
 limit. We tried to correlate the cosmological BAU from resonant leptogenesis  with the low energy observables, notably, NDBD
 and LFV with a view to finding a common parameter space where they coexists.
\end{abstract}
\maketitle

\section{INTRODUCTION}

 The landmark discovery of neutrino flavor oscillations from neutrino experiments like MINOS \cite{minos}, T2K \cite{t2k}, Double Chooz \cite{doublechooz}, 
 Daya Bay \cite{dayabay}, RENO \cite{reno} etc and hence the evidence of neutrino mass and mixing have immense impact on our perception of the dynamics
 of the universe. Regardless of its enormous success, the Standard Model (SM) of particle physics is considered an insufficient theory, owing to the fact 
 that it fails to address some of the vital questions like, the origin of the tiny neutrino mass, Baryon Asymmetry of the universe (BAU), Dark matter (DM),
 Lepton Number Violation (LNV), Lepton Flavor Violation (LFV) and various other cosmological problems.

\par There are many beyond standard model (BSM) frameworks to realize these observables. Amongst them, the seesaw mechanism is the simplest way to understand the smallness of 
neutrino masses, which is further categorized into type I, type II, type III, Inverse seesaw (SS) mechanisms \cite{type1}\cite{type2}\cite{type3}\cite{inverse}. In type I seesaw,
the introduction of SM gauge singlet RH neutrinos,
gives rise to the light neutrino mass matrix of the form, $ M_\nu\approx -M_DM^{-1}_{RR}M^T_D$, with a heavy-light neutrino mixing of order $M_D M^{-1}_{RR}$, where,$ M_D$ and $M_{RR}$ are
the Dirac and Majorana masses respectively.
Notwithstanding, one of the most appealing frameworks BSM, in which the seesaw mechanisms arises naturally is the left-right symmetric model (LRSM) which is based on the gauge group, 
$ \rm SU(3)_c\times SU(2)_L\times SU(2)_R\times U(1)_{B-L}$. Here, the RH neutrinos are a necessary part of the model, which acquires a Majorana mass when the $SU(2)_R$ 
symmetry is broken at a scale $v_R$. This is quite analogous to the way in which the charged fermions get masses in the SM by Higgs mechanism when $SU(2)_L$  gauge symmetry
 is broken at a scale $v$. \par The RH neutrinos which exist in the seesaw mechanism, besides explaining the neutrino flavour oscillation and neutrino mass can also throw light
 on one of the most enthralling problems of particle physics and cosmology, the matter-antimatter asymmetry of the universe, i.e, excess of baryons over anti baryons in the 
 universe. The decay of the lightest right handed neutrino, $N_1$ can naturally give rise to an excess of baryons over anti baryons in the universe consistent with the cosmological 
observable constrained by Big bang Nucleosynthesis and determined recently with a good precision by WMAP experiment as,
\begin{equation}
 \eta_B=\frac{n_B}{n_\gamma}=\left(6.5^{+0.4}_{-0.3}\right)\times 10^{-10}.
\end{equation}

The decay of $N_1$ can satisfy all the three Sakharov conditions \cite{sakharov} as required for successful generation of  $\eta_B$ as there is sufficient CP and C violation, there is Baryon
number violation and can also occur out of thermal equillibrium. TeV scale LRSM provides an alluring 
class of SS models which can be probed at LHC. Matter Antimatter asymmetry is now generated by a resonant baryogenesis mechanism with atleast two Quasi Degenerate RH neutrinos
in TeV range with a mass difference comparable to their decay widths \cite{decaywidth}. The TeV scale new particles in LRSM also leads to interesting collider signals.

\par The possible observation of NDBD would play an important role in understanding the origin of BAU as it would imply that lepton number indeed is not conserved
(one of the essential conditions for leptogenesis \cite{leptogenesis}). Furthermore, the Majorana nature \cite{maj} of neutrinos would also be established from NDBD. The latest experiments 
\cite{ndbd} that have improved the lower bound of the half life of the decay process include KamLAND-Zen \cite{kamland} and GERDA \cite{gerda} which uses Xenon-136 and 
Germanium-76 nuclei respectively. Incorporating the results from first and second phase of the experiment, KamLAND-Zen imposes the best lower limit on the decay half life using Xe-136 as
$ \rm T_{1/2}^{0\nu}>1.07\times 10^{26}$ yr at $ 90\%$ CL and the corresponding upper limit of effective Majorana mass in the range (0.061-0.165)eV.

The observation of CP violation in
lepton sector, in neutrino osillation experiment and NDBD would suggest the existence of CP violation at high energy which might be related to the one responsible for leptogenesis. 
The observation of LNV in NDBD and in addition possibly of CP violation in lepton sector would be a strong indication of leptogenesis as an explaination of baryon asymmetry. It would be interesting to
explore the existence of CP violation in leptonic sector due to Majorana CP phases in the light of leptogenesis. 
\par 

\par Another important issue of discussion in collider is the relative values of mass of the gauge bosons and heavy right handed neutrinos. However there are theoretical
arguments based on vacuum stability which suggests that the heavy neutrinos are lighter than the RH gauge bosons that appears in the LRSM for a large parameter space. 
 Again, it has been pointed out in literature that to account for a successful leptogenesis in TeV scale LRSM, the mass of the RH gauge boson, 
$M_{W_R}$ has to be larger than the value obtainable at the LHCs. They have found a lower bound of 18 TeV for successful leptogenesis from the decay of heavy RH neutrino
with maximum CP asymmetry, $\varepsilon=1$. \cite{18tev}. This result is much significant as it can provide a way to falsify leptogenesis, if mass of a gauge 
boson below below this limit is found in experiments. From the significant outcome of the work, \cite{18tev}, the authors of \cite{3tev} have shown that for 
specific symmetry textures of $M_D$ and $M_{RR}$ in the seesaw formula and by considering larger Yukawa couplings, the bound for leptogenesis can be largely 
weaker, i.e. $M_{W_R}>3$ TeV and $M_N \leq M_{W_R}$ which is possible owing to the sizeable reduction of dilution effects from $W_R$ mediated decays and
scatterings. They have again reanalyzed their work \cite{3tev} in \cite{10tev} and came out with a lower bound of $M_{W_R} >10$ TeV for successful leptogenesis
in a generic LRSM with large light- heavy mixing. The consistency has also been pointed out for other low energy constraints like NDBD, LFV etc.

\par In LRSM, there are several contributions to NDBD that involve left and right handed sectors individually as well as others that involve both sectors through left-right mixing
accompanied by both light and heavy neutrinos.
Left-right mixing is always a ratio of the Dirac and Majorana mass scales $(M_DM^{-1}_{RR})$ which appears in the type I seesaw formula. NDBD involving left-right mixing can be 
enhanced for specific Dirac matrices. 
For large left right mixing, significant contributions to NDBD arises from the mixed diagrams with simultaneous mediation of $W_L$ and $W_R$ accompanied by light left handed neutrino
and heavy right handed neutrinos, known as $\lambda$ and $\eta$ contribtions to NDBD , although the later is a bit suppressed by the mixing between left and right
handed gauge bosons. It has been studied in many of the earlier works in the framework of LRSM (see ref.\cite{ls1}\cite{dbd1}\cite{ndbddb2}) The other new physics contributions are also suppressed for a larger
gauge boson mass, $M_{W_R}>$10 TeV which gives sizeable baryogenesis. Furthermore,
the LFV processes are seeking great interest in recent times as the experiments to detect them are becoming increasingly precise. The decay processes, $\left(\mu\rightarrow 3e\right)$ and 
$\left(\mu\rightarrow e\gamma\right)$ are simplest to detect with the current experimental limits for these low energy processes as $<1.0\times 10^{-12}$ and $<4.2\times 10^{-13}$ respectively.

\par Apart from the new physics contributions to NDBD in LRSM as available in literature, it is important to study the linkage between 
baryogenesis and other low scale phenomenon like NDBD, LFV etc. In this context, with the previous results aforementioned in mind \cite{18tev}\cite{3tev}\cite{10tev}\cite{ls1}\cite{dbd1}\cite{ndbddb2} we have done a phenomenological study of 
leptogenesis in TeV scale LRSM by considering different values of RH gauge boson mass within and above
the current collider limits. In particular we have considered the $SU(2)_R$ breaking scale to be 5 TeV, 10 TeV and 18 TeV 
(the bounds as available in literature) in order to check the consistency of the results and thereby tried to link baryogenesis with NDBD for these particular values of
gauge boson mass.
Again regarding the $\lambda$ and $\eta$ contributions to be valid, we need to have a large
left-right mixing. But for a generic TeV scale seesaw model, without considering any particular structure for the Dirac and Majorana masses, in order to account for 
neutrino mass of the order of sub eV, keeping the heavy masses of TeV scale, the Dirac mass is of the order of MeV. This leads to a not so large left-right mixing parameter,
$\zeta\approx 10^{-6}$.  Since we have seen non negligible effects of the momentum dependent mechanisms in NDBD for not so large left light mixing, 
we studied all the possible contributions to NDBD. To co-relate with baryogenesis, we have considered only the momentum dependent mechanisms of NDBD, i.e., the $\lambda$ and $\eta$ contributions
to NDBD due to light-heavy and gauge boson mixing. Since the effective mass govering NDBD is dependent upon
the Majorana phases, $\alpha$ and $\beta$, it would be compelling to examine if there exist a link between NDBD and BAU. Besides, the study of LFV processes will also
provide insights about the mechanism of NDBD. LRSM at the TeV scale interlinks high energy collider physics to the low energy observables like NDBD and other LFV processes.
So we tried to correlate all these high and low energy phenomenon and find out if there exist a common parameter space accessible at colliders where leptogenesis can be simultaneously
realized.
\par This paper is outlined as follows. In the next section, we present the left-right
symmetric model framework with its particle contents and the origin of neutrino mass. In section \ref{sec:level4}, we summarized the implications of TeV scale LRSM in processes like BAU and other
low energy observables like NDBD, LFV. In section \ref{sec:level5}, we present our numerical 
 analysis  and results and then give our conclusion  in section \ref{sec:level6}

\section{LEFT RIGHT SYMMETRIC MODEL(LRSM) AND NEUTRINO MASS}{\label{sec:level3}}

In the generic LRSM \cite{LRSM}, the fermions are assigned to the gauge group  $ \rm SU(3)_c\times SU(2)_L\times SU(2)_R\times U(1)_{B-L}$ \cite{genericlrsm} \cite{LRSM} which
is a very simple extension of the standard model gauge group, $ \rm SU(3)_c\times SU(2)_L\times U(1)_Y$, that provides a UV complete
seesaw model where the type I and II seesaw arises naturally. Most of the problems 
like parity violation of weak interaction, masssless neutrinos, CP problems, hierarchy problems etc can be realized in the framework of LRSM. The seesaw scale is
identified as the breaking of the $SU(2)_R$ symmetry.
In this model, the electric charge takes a form,$ Q=T_{3L}+T_{3R}+\frac{B-L}{2}$ \cite{Q}, where $ \rm T_{3L}$ and $ \rm T_{3R}$ are the 3rd components of isospin under $ \rm SU(2)_L$ and $ \rm SU(2)_R$. In LRSM, the left and right handed components of the fields are treated
on the same footing. The leptons (LH and RH) that transform in L-R symmetric gauge group are assigned with quantum numbers $(1,2,1,-1)$ and $(1,1,2,-1)$ respectively under
$ \rm SU(3)_c\times SU(2)_L\times SU(2)_R\times U(1)_{B-L}$. The Higgs sector in LRSM consists of two scalar triplets, $\Delta_L(1,2,1,-1)$, $\Delta_R(1,1,2,-1)$
and  a Bidoublet with quantum number $ \rm \phi(1,2,2,0)$.
A 2$\times$ 2 matrix representation for the Higgs bidoublets and the $ \rm SU(2)_{L,R}$ triplets is given as,

\begin{equation}\label{eqx3}
\phi=\left[\begin{array}{cc}
             \phi_1^0 & \phi_1^+\\
             \phi_2^- & \phi_2^0
            \end{array}\right]\equiv \left( \phi_1,\widetilde{\phi_2}\right), \Delta_{L,R}=\left[\begin{array}{cc}
                     {\delta_\frac{L,R}{\sqrt{2}}}^+ & \delta_{L,R}^{++}\\
                     \delta_{L,R}^0 & -{\delta_\frac{L,R}{\sqrt{2}}}^+ .
                    \end{array}\right].
\end{equation}

The VEVs  of the neutral component of the Higgs field are $ \rm v_R,v_L,k_1,k_2$ respectively.The VEV $ \rm v_R$ breaks the $ \rm SU(2)_R$ symmetry and sets the mass scale
for the extra gauge bosons $ \rm (W_R$ and Z$ \rm \ensuremath{'})$ and for right handed neutrino
field $ \rm (\nu_R)$. The VEVs $ \rm k_1$ and $ \rm k_2$ serves the twin purpose of breaking the remaining the  $ \rm SU(2)_L\times U(1)_{B-L}$ symmetry down to
$ \rm U(1)_{em}$, thereby setting 
the mass scales for the observed $ \rm W_L$ and Z bosons and providing Dirac masses for the quarks and leptons. Clearly, $ \rm v_R$ must be significantly larger 
than $ \rm k_1$ and $ \rm k_2$ in order for $ \rm W_R$ and Z $\ensuremath{'}$ to have greater masses than the $W_L$ and Z bosons. $v_L$ is the VEV of $\Delta_L$, 
it plays a significant role 
in the seesaw relation which is the characteristics of the LR model and can be written as,

\begin{equation}\label{eqx7}
 <\Delta_L>=v_L=\frac{\gamma k^2}{v_R}.
\end{equation}

\par The Yukawa lagrangian in the lepton sector is given by,
\begin{equation}\label{eqx8}
 \mathcal{L}=h_{ij}\overline{\Psi}_{L,i}\phi\Psi_{R,j}+\widetilde{h_{ij}}\overline{\Psi}_{L,i}\widetilde{\phi}\Psi_{R,j}+f_{L,ij}{\Psi_{L,i}}^TCi\sigma_2\Delta_L\Psi_{L,j}+f_{R,ij}{\Psi_{R,i}}^TCi\sigma_2\Delta_R\Psi_{R,j}+h.c.
\end{equation}

Where the family indices i,j are summed over, the indices $i,j=1,2,3$ represents the three generations of fermions. $C=i\gamma_2\gamma_0$ is the charge conjugation 
operator, $\widetilde{\phi}=\tau_2\phi^*\tau_2$ and $\gamma_{\mu}$ are the Dirac matrices. Considering discrete parity symmetry, the Majorana Yukawa couplings $f_L=f_R$ (for left-right symmetry) gives rises
to Majorana neutrino mass after electroweak symmetry breaking  when the triplet Higgs $\Delta_L$ and $\Delta_R$ acquires non zero vacuum expectation value.
Then equation (\ref{eqx8}) leads to $6\times6$ neutrino mass matrix as shown in reference 2 of \cite{ls1} 

\begin{equation}\label{eqx8}
M_\nu=\left[\begin{array}{cc}
              M_{LL}&M_D\\
              {M_D}^T&M_{RR}
             \end{array}\right],
\end{equation}

where 
\begin{equation}\label{eqx9}
M_D=\frac{1}{\sqrt{2}}(k_1h+k_2\widetilde{h}), M_{LL}=\sqrt{2}v_Lf_L, M_{RR}=\sqrt{2}v_Rf_R,
\end{equation}

where $M_D$, $M_{LL}$ and $M_{RR}$ are the Dirac neutrino mass matrix, left handed and right handed mass matrix respectively. Assuming $M_L\ll M_D\ll M_R$, the 
light neutrino mass, generated within a type I+II seesaw can be written as,

\begin{equation}\label{eqx10}
 M_\nu= {M_\nu}^{I}+{M_\nu}^{II},
\end{equation}

\begin{equation}\label{eqx11}
 M_\nu=M_{LL}+M_D{M_{RR}}^{-1}{M_D}^T
      =\sqrt{2}v_Lf_L+\frac{k^2}{\sqrt{2}v_R}h_D{f_R}^{-1}{h_D}^T,
\end{equation}

where the first and second terms in equation (\ref{eqx11}) corresponds to type II seesaw and type I seesaw mediated by RH neutrino respectively.
Here,
\begin{equation}\label{eqx12}
 h_D=\frac{(k_1h+k_2\widetilde{h})}{\sqrt{2}k} , k=\sqrt{\left|{k_1}\right|^2+\left|{k_2}\right|^2}.
\end{equation}

In the context of LRSM both type I and type II seesaw terms can be written in terms of $M_{RR}$ which arises naturally at a high energy scale as a result
of spontaneous parity breaking. In LRSM the Majorana Yukawa couplings $f_L$ and $f_R$ are same (i.e, $f_L=f_R$) and the VEV for left handed triplet $v_L$ can be written as,

\begin{equation}\label{eqx13}
v_L=\frac{\gamma {M_W}^2}{v_R}.
\end{equation}

Thus equation (\ref{eqx11}) can be written as ,

\begin{equation}\label{eqx14}
M_\nu=\gamma(\frac{M_W}{v_R})^2M_{RR}+M_D{M_{RR}}^{-1}{M_D}^T.
\end{equation}

In literature, (reference \cite{breaking} \cite{ndbddb2}) author define the dimensionless parameter $\gamma$ as,

\begin{equation}\label{eqx15}
 \gamma=\frac{\beta_1k_1k_2+\beta_2{k_1}^2+\beta_3{k_2}^2}{(2\rho_1-\rho_3)k^2}.
\end{equation}

Here the terms $\beta$, $\rho$ are the dimensionless parameters that appears in the expression of the Higgs potential.
\par Again, the neutrino mass matrix as given in \ref{eqx8} can be digonalized by a $6\times 6$ unitary matrix, as follows,
\begin{equation}\label{eqx16}
\mathcal{V}^T M_\nu \mathcal{V}=\left[\begin{array}{cc}
              \widehat{M_\nu}&0\\
              0&\widehat{M}_{RR}
             \end{array}\right],
\end{equation}
where, $\mathcal{V}$ represents the diagonalizing matrix of the full neutrino mass matrix, $M_\nu$, $\widehat{M_\nu}= diag(m_1,m_2,m_3)$, with $m_i$ being the light neutrino masses and 
$\widehat{M}_{RR}= diag(M_1,M_2,M_3)$, with $M_i$ being the heavy RH neutrino masses. The diagonalizing matrix is represented as,
\begin{equation}\label{eqx17}
 \mathcal{V}=\left[\begin{array}{cc}
              U&S\\
              T&V
             \end{array}\right] \approx \left[\begin{array}{cc}
              1-\frac{1}{2}RR^{\dagger}&R\\
              -R^\dagger&1-\frac{1}{2}R^\dagger R
             \end{array}\right] \left[\begin{array}{cc}
              V_\nu&0\\
              0&V_R
             \end{array}\right],
\end{equation}

where, R describes the left-right mixing and given by,

\begin{equation}\label{eqx18}
 R=M_D M^{-1}_{RR}+\mathcal{O} (M^3_D{(M^{-1}_{RR})}^3).
\end{equation}
The matrices U, V, S and T are as follows,
\begin{equation}\label{eqx19}
 U=\left[1-\frac{1}{2}M_DM^{-1}_{RR}{(M_DM^{-1}_{RR})}^\dagger\right]V_\nu,
 \end{equation}
 \begin{equation}\label{eqx20}
 V=\left[1-\frac{1}{2}{(M_DM^{-1}_{RR})}^\dagger M_DM^{-1}_{RR}\right]V_R,
 \end{equation}
 \begin{equation}\label{eqx21}
 S=M_DM^{-1}_{RR}v_Rf_R,
 \end{equation}
 \begin{equation}\label{eqx22}
 T=-(M_DM^{-1}_{RR})^\dagger V_\nu.
\end{equation}

The leptonic charged current interaction in flavour basis is given by,
\begin{equation}
\mathcal{L}^{lepton}_{CC}=\frac{g}{\sqrt{2}}\left[\overline{l^{'}}\gamma^\mu P_L \nu^{'}W^{-}_{L_\mu}+\overline{l^{'}}\gamma^\mu P_R \nu^{'}W^{-}_{R_\mu}\right]+h.c,
\end{equation}
where,
\begin{equation}\label{eqx23}
\left[\begin{array}{cc}
              W^{\pm}_L\\
               W^{\pm}_R
             \end{array}\right]= \left[\begin{array}{cc}
             \cos \zeta & \sin\zeta e^{i\alpha}\\
              -\sin \zeta e^{-i\alpha} & \cos\zeta
             \end{array}\right]\left[\begin{array}{cc}
               W^{\pm}_1\\
               W^{\pm}_2
             \end{array}\right] ,
\end{equation}

characterises the mixing between L-R gauge bosons with,

\begin{equation}\label{eqx24}
 \tan 2\zeta= -\frac{2 k_1 k_2}{v^2_R-v^2_L}.
\end{equation}
With negligible mixing, the gauge boson masses become,
\begin{equation}\label{eqx25}
 M_{W_L}\approx M_{W_1}\approx \frac{g}{2}k_+, M_{W_R}\approx M_{W_2}\approx \frac{g}{\sqrt{2}}v_R .
\end{equation}.
Assuming $k_2< k_1\Rightarrow \zeta \approx -\frac{k_1k_2}{v^2_R} \approx -2\frac{k_2}{k_1}{\left(\frac{M_{W_L}}{M_{W_R}}\right)}^2$. T and S in equation \ref{eqx17} describes
the left-right mixing and can be written as $\frac{L}{R}$, gauge boson mixing angle $\zeta$ is of order ${\left(\frac{L}{R}\right)}^2$.

\section{Resonant Leptogenesis, NDBD and LFV in TeV scale LRSM}{\label{sec:level4}}

As illustrated in several earlier works, for TeV scale seesaw models, a simple approach for generating adequate lepton asymmetry is to use resonant leptogenesis
(RL) \cite{RL}, which craves for at least two heavy RH Majorana neutrinos to be nearly degenerate, which we have already considered in our analysis. With Quasi degenerate
RH neutrino masses for at least two RH neutrinos, BAU/leptogenesis can be efficient at lower mass scales, but for this case generally a specific
flavour structure is generally considered which allows for large Yukawa couplings which serves the twin purpose of leptogenesis to be efficient as well as 
it can be tested in experiments. Nevertheless, as far as Dirac neutrino mass matrix is concerned,
we have not considered any particular structure of the matrix but a general form which is obtained from the type I seesaw when the Majorana mass matrix and the light neutrino mass
matrix is considered to be known. The neutrino mass matrices is such that it fits the current neutrino oscillation data. The basic focus of our work is to relate the 
lepton asymmetry with the low observable phenomenons like NDBD, rather than only BAU and NDBD or LFV and to find a common parameter space where all them them holds true.

In the framework of TeV scale LRSM, the presence of the RH neutrinos (type I SS) and the scalar triplets (type II SS) suggests their decays which give rise to lepton asymmetry.
However we will only consider the decay of the heavy RH neutrinos for generating lepton asymmetry . The decay of the scalar triplet $\Delta_L$ would not much affect on our result as above TeV 
scale, decay of RH neutrinos are in thermal equillibrium and hence they would wash out any kind of preexisting lepton asymmetry and so we have ignored it \cite{10tev}. So the dominant contribution would come from
the type I seesaw term.

The two heavy RH Majorana neutrinos decay via the decay modes,  $N_i\rightarrow l+\phi^{c} $ and its 
CP conjugate process,  $N_i\rightarrow l^c+\phi $ which can occur at both tree and one loop levels. Hence, their CP violating asymmetry $\epsilon_i$ 
which arises from the interference between the tree level amplitude and its self-energy \cite{selfenergy} correction is defined as \cite{zing},

\begin{equation}\label{eqa7}
 \epsilon_i= \frac{\Gamma\left(N_i\rightarrow l+\phi^{c}\right)-\Gamma\left(N_i\rightarrow l^c+\phi\right)}{\Gamma\left(N_i\rightarrow l+\phi^{c}\right)+\Gamma\left(N_i\rightarrow l^c+\phi\right)}.
\end{equation}
 The decay rates of the  heavy neutrino decay processes are governed 
by the Yukawa couplings, and is given by,
\begin{equation}\label{eqa8}
 \Gamma_i= {\left(Y^\dagger_\nu Y_\nu\right)}_{ii}\frac{M_i}{8\pi}.
\end{equation}
 
An essential condition for RL is that the mass difference between the two heavy RH neutrinos must be comparable to the decay width ( i.e.,$ M_i-M_j \approx \Gamma$). In this
case, the CP aymmetry becomes very large (even of order 1). The  CP violating asymmetry $\epsilon_i$ is thus given by,

 \begin{equation}\label{eqa9}
 \epsilon_i= \frac{Im\left[{\left(Y^\dagger_\nu Y_\nu\right)}^2_{ij}\right]}{\left(Y^\dagger_\nu Y_\nu\right)_{11}\left(Y^\dagger_\nu Y_\nu\right)_{22}}.
 \frac{\left(M^2_i-M^2_j\right)M_i \Gamma_j}{\left(M^2_i-M^2_j\right)+M^2_i \Gamma^2_j},
\end{equation}

where,
\begin{equation}
 \frac{Im\left[{\left(Y^\dagger_\nu Y_\nu\right)}^2_{ij}\right]}{\left(Y^\dagger_\nu Y_\nu\right)_{11}\left(Y^\dagger_\nu Y_\nu\right)_{22}}\approx 1.
\end{equation}

The variables i, j run over 1 and 2, $i \neq j$.
\par The CP violating asymmetries $\epsilon_1$ and $\epsilon_2$ can give rise to a net lepton number asymmetry, provided the expansion rate of the universe is larger than
$\Gamma_1$ and $\Gamma_2$. This can further be partially converted into baryon asymmetry of the universe by B+L violating sphaleron \cite{sphaleron} processes.

Now that there are several new heavy particles in LRSM, many new physics contributions
to NDBD arises in addition to the standard contribution.  It has been extensively studied in many of the earlier works (see ref. \cite{dbd1}\cite{ndbddb2}).
Amongst the new physics contributions to $0\nu\beta\beta$ decay, notable are the contributions coming from the exchange of the heavy gauge bosons ( $ {W_L}^-$  and  $ {W_R}^-$ ), the 
both the left  and right handed gauge bosons (mixed diagrams, $\lambda$ and $\eta$) as well the scalar triplet ($\Delta_L$  and $\Delta_R$ ) contributions. The amplitude of 
these processes mostly depends upon the mixing between light and heavy neutrinos, the  leptonic mixing matrix elements, the mass of the heavy neutrino ($M_i$), the mass of the gauge
bosons, ${W_L}^-$ and ${W_R}^-$ , the mass of the triplet Higgs  as well as their  coupling to leptons, $f_L$ and $f_R$ .

\par However in our present work, we have considered only three of the aforesaid contributions to NDBD. The ones mediated by ${W_R}^-$ and the momentum dependent mechanisms,
i.e., the contributions to NDBD from $\lambda$ and $\eta$ diagrams which involves the light and heavy
neutrino mixings and the mixing between ${W_L}^-$ and ${W_R}^-$ bosons (considering a small light heavy neutrino mixing of $\mathcal{O}$($10^{-6}$)  .
The amplitudes of the contributions are given in several earlier works like \cite{ndbddb2}.
The mass scales for the heavy particles has been assumed to be $\approx TeV $, with $M_{W_R}> M_N$. Under these assumptions, the amplitude  for the 
light-heavy mixing contribution which is proportional to $ \frac{{m_D}^2}{M_R}$ remains very small (since $m_\nu \approx \frac{{m_D}^2}{M_R} \approx (0.01-0.1) eV$, $m_D \approx
(10^5-10^6)$ eV which implies $ \frac{m_D}{M_R} \approx (10^{-7}-10^{-6})$ eV).

Again, the contribution
from ${\Delta_L}^-$, ${W_L}^-$ is suppressed by the type II seesaw contribution to light neutrino mass and hence neglected here. Considering these contributions
we have studied the NDBD and tried to correlate the effective mass governing the process with the BAU for different gauge boson masses in 
TeV scale LRSM.

As has been pointed out that successful low scale RL requires an absolute lower bound of 18 TeV on the mass of the RH gauge boson and recent work predicted that it can be
produced for considerably lower value of $M_{W_R}$ accessible at LHCs considering relatively large Yukawa couplings. Again, although it has been illustrated as the
light-heavy neutrino mixing to be sufficiently large in TeV scale LRSM inorder to get dominant NDBD contributions from the momentum dependent mixed diagrams, $\lambda$ 
and $\eta$, we have seen that a sizeable amount of BAU  and effective mass governing NDBD ( from $\lambda$ and $\eta$ diagrams) consistent with the experimental value is observed
by considering a general structure of the Dirac mass matrix and not so large light-heavy neutrino mixing parameter. Without considering any special structure of $M_D$ and $M_{RR}$
in generic TeV scale LRSM, inorder to get light neutrino mass of the order of sub eV, $M_D$ has to be fine tuned to be very small which results in a lower value of the light heavy 
neutrino mixing parameter, $\zeta$. But, in our present work, by considering
a smaller $\zeta$ value, we have tried to correlate the effective mass from purely RH contribution and the suppressed effective mass coming from $\lambda$ and $\eta$ 
conributions with leptogenesis at a TEV scale LRSM.

\par The heavy Majorana neutrinos that takes part in explaining BAU as well as NDBD also plays a significant role in giving rise to experimentally testable rates of LFV
processes like, $\mu\rightarrow e\gamma$, $\mu\rightarrow 3e$, $\mu\rightarrow e$ etc. The different neutrino Yukawa couplings for each lepton flavour have a considerable 
impact on leptogenesis with nearly degenerate heavy neutrino mass. Owing to the presence of some new heavy particles in the LRSM, the LFV proceses are mediated by these
heavy neutrinos and doubly charged triplet Higgs bosons.

 The relevant BR for the
process $(\mu\rightarrow 3e )$ is defined as, \cite{lfv1}
\begin{equation}\label{eq43}
 BR\left(\mu\rightarrow 3e\right)=\frac{1}{2}{\left|h_{\mu e}h_{ee}^{*}\right|}^{2}\left(\frac{{m_{W_L}}^4}{{M_{\Delta_L}^{++}}^4}+\frac{{m_{W_R}}^4}{{M_{\Delta_R}^{++}}^4}\right).
\end{equation}
 Where $h_{ij}$ describes the lepton Higgs coupling in LRSM and is given by,

 \begin{equation}\label{eq44}
  h_{ij}=\sum_{n=1}^{3}V_{in}V_{jn}\left(\frac{M_n}{M_{W_R}}\right), i,  j=e,\mu,\tau.
 \end{equation}
  \par For $\mu\rightarrow e\gamma$, the BR is given by, \cite{lfv1}
  \begin{equation}\label{eqa}
   BR\left(\mu\rightarrow e\gamma\right)= 1.5\times 10^{-7}{\left|g_{lfv}\right|}^2{\left(\frac{1 TeV}{M_{W_R}}\right)}^4,
   \end{equation}\\
  where, $ g_{lfv}$ is defined as,
  \begin{equation}\label{eqb}
   g_{lfv}=\sum_{n=1}^{3}V_{\mu n}{V_{e n}}^*{\left(\frac{M_n}{M_{W_R}}\right)}^2=\frac{\left[M_R {M_R}^*\right]_{\mu e}}{{M_{W_R}}^2}.
  \end{equation}

\par The sum is over the heavy neutrinos only. $ \rm M_{\Delta_{L,R}}^{++}$ are the masses of the doubly charged bosons, $ \rm {\Delta_{L,R}}^{++}$, V is the mixing matrix
 of the right handed neutrinos with the electrons and muons. $ \rm M_n(n=1,2,3) $ are the right handed neutrino masses.

\par Several new sources of LFV are present in new physics BSM in LRSM due to the additional RH current interactions, which could lead to considerble
LFV rates for TeV scale $v_R$. LFV in the LRSM has been studied in many previous works. There are various LFV processes providing constraints on the masses
of the right handed neutrinos and doubly charged scalars. It turns out that  the process $\mu\rightarrow 3e $ induced by doubly charged bosons
$\Delta_L^{++}$ and $\Delta_R^{++}$  and $\mu\rightarrow e\gamma$ provides the most relevant constraint.  
The upper limits of branching ratio of the process  $\mu\rightarrow 3e $ is $<1.0\times 10^{-12}$  \cite{SINDRUM} at $ 90\%$ CL was obtained  by the 
SINDRUM experiment. Furthermore, the Mu3e collaboration has also submitted a letter of intent to PSI to perform a new  improved search for the decay
$\mu\rightarrow 3e $ with a sensitivity of $10^{-16}$  at $95\%$ CL \cite{sindrum2} which corresponds to an improvement by four orders of magnitude 
compared to the former SINDRUM experiment. While for the LFV process, $\mu\rightarrow e\gamma$, the BR is established to be
$<4.2\times 10^{-13}$  \cite{muegamma} at $ 90\%$ CL by the MEG collaboration. Considering these
contributions from heavy righthanded neutrinos and Higgs scalars, the expected branching ratios and conversion rates of the above processes have been 
calculated in the LRSM in the work (first reference in \cite{lfv}).

\section{NUMERICAL ANALYSIS AND RESULTS}{\label{sec:level5}}
 With reference to several earlier works \cite{18tev}\cite{3tev}\cite{10tev}\cite{ndbddb2}\cite{baundbd} for TeV scale LRSM, we carried out an extensive  study for 
 RL, NDBD and LFV, with a view to finding a common parameter space for these observabes. It is reasonable to check if the mass matrices that can explain the BAU of the universe can also provide 
 sufficient parameter space for other low energy observables like NDBD, LFV etc. For NDBD, we have considered the mixed LH-RH contribution along with the purely RH neutrino
 contribution, considering a generalized structure for the Dirac mass matrix. The Dirac and Majorana mass matrices in our case are determined using 
 the type I seesaw formula ( as shown in appendix) and type II seesaw (equation\ref{eqa5}) respectively  which satisfies the recent neutrino oscillation data. Whereas, in the previous works, the authors have considered
 specific Dirac and Majorana textures resulting in light neutrinos via type I seesaw with large light heavy neutrino mixing. They have chosen large Yukawa 
 couplings as allowed by specific textures for calculation of the lepton asymmetry. As stated in \cite{10tev}, we have been found that it is possible to  observe BAU with 
 a lower $W_R$ mass, in our case it is 5 TeV. Further, we have also correlated the LFV of the process, $\mu\rightarrow 3e $, $\mu\rightarrow e\gamma $ and  with lightest neutrino mass and atmospheric mixing angle. 
In this section we present a detailed analysis of our work by dividing it into several subsections, firstly BAU, then NDBD and then LFV. 
\subsection{Baryogenesis via Leptogenesis}

The formula for light $\nu$ masses in LRSM can be written as,
\begin{equation}\label{eqa1}
 M_\nu={ M_\nu}^{I}+{ M_\nu}^{II},
\end{equation}
where the type I seesaw mass term is,
\begin{equation}\label{eqa2}
{ M_\nu}^{I}= M_D{M_{RR}}^{-1}{M_D}^T.
\end{equation}
We have considered a tribimaximal mixing (TBM) pattern, such that,
\begin{equation}\label{eqa3}
 { M_\nu}^{I}= U_{(TBM)}U_{Maj}{M_\nu}^{I(diag)}{U_{Maj}}^T{U_{(TBM)}}^T,
\end{equation}
where $ \rm {M_\nu}^{I(diag)}=X{M_\nu}^{(diag)}$ \cite{X}, 
the parameter X is introduced to describe the relative strength of the type I and II seesaw terms. It can take any numerical value
provided the two seesaw terms gives rise to correct light neutrino mass matrix. In our case, we have considered X=0.5 \cite{X}, i.e., equal 
contributions from both the seesaw terms.
Thus, equation (\ref{eqa1}) can be written as,
\begin{equation}\label{eqa4}
 U_{PMNS}{M_\nu}^{(diag)} {U_{PMNS}}^T={ M_\nu}^{II}+U_{(TBM)}U_{Maj}X{M_\nu}^{(diag)}{U_{Maj}}^T{U_{(TBM)}}^T,
\end{equation}
 where, $\rm U_{PMNS}$ is the diagonalizing matrix of the light neutrino mass matrix, $M_\nu$  and is given by,

\begin{equation}\label{eq5}
\rm U_{PMNS}=\left[\begin{array}{ccc}
c_{12}c_{13}&s_{12}c_{13}&s_{13}e^{-i\delta}\\
-c_{23}s_{12}-s_{23}s_{13}c_{12}e^{i\delta}&-c_{23}c_{12}-s_{23}s_{13}s_{12}e^{i\delta}&s_{23}c_{13}\\
 s_{23}s_{12}-c_{23}s_{13}c_{12}e^{i\delta}&-s_{23}c_{12}-c_{23}s_{13}s_{12}e^{i\delta}&c_{23}c_{13}
\end{array}\right]U_{Maj}.
\end{equation}

The abbreviations used are $c_{ij}$= $\cos\theta_{ij}$, $s_{ij}$=$\sin\theta_{ij}$, $\delta$ is the Dirac CP phase while the diagonal phase matrix,
$ \rm U_{Maj}$ is $ \rm diag (1,e^{i\alpha},e^{i(\beta+\delta)}) $, contains the Majorana phases $ \rm \alpha$ and $ \rm \beta$. We have adopted the recent neutrino oscillation
data in our analysis as in the table \ref{t1},
\begin{table}[h!]
\centering
\begin{tabular}{||c| c| c||}
\hline
PARAMETERS & $3 \sigma$ RANGES & BEST FIT$\pm 1 \sigma$\\ \hline
$\Delta m_{21}^2[10^-5 \rm eV^2]$ & 6.93-7.97 & 7.37\\ \hline
$\Delta m_{31}^2[10^-3  \rm eV^2]$(NH) & 2.37-2.63 & 2.50\\
$\Delta m_{23}^2[10^-3 \rm eV^2]$(IH) & 2.33-2.60& 2.46\\ \hline
$\sin^2{\theta_{12}}$ & 0.250-0.354 & 0.297\\ \hline
$\sin^2{\theta_{23}}$(NH) & 0.379-0.616 & 0.437\\ 
(IH) & 0.383-0.637 & 0.569\\ \hline
$\sin^2{\theta_{13}}$(NH) & 0.0185-0.0246 & 0.0214\\ 
(IH) & 0.0186-0.0248 &  0.0218\\ \hline
$\delta/\pi$ & 0-2(NH)& 1.35\\
         & 0-2(IH)& 1.32\\ \hline
\end{tabular}
\caption{Global fit 3$\sigma$ values of $\nu$ oscillation parameters \cite{sigma}} \label{t1}
\end{table}

 From type II seesaw mass term,$ \rm M_{RR}$ can be written in the form(from reference \cite{newphysics})as,
\begin{equation}\label{eqa5}
 M_{RR}=\frac{1}{\gamma}{\left(\frac{v_R}{M_{W_L}}\right)}^2{ M_\nu}^{II},
\end{equation}
\begin{equation}\label{eqa6}
 U_{R}{M_{RR}}^{(diag)}{U_{R}}^T=\frac{1}{\gamma}{\left(\frac{v_R}{M_{W_L}}\right)}^2{ M_\nu}^{II},
\end{equation}
\begin{equation}
 { M_\nu}^{II}= U_{PMNS}{M_\nu}^{(diag)} {U_{PMNS}}^T- U_{(TBM)}U_{Maj}X{M_\nu}^{(diag)}{U_{Maj}}^T{U_{(TBM)}}^T.
\end{equation}

\begin{figure}[h!]
\includegraphics[width=0.46\textwidth]{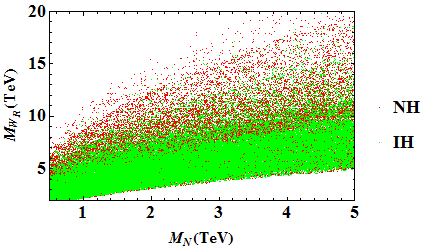}
\caption{$M_{W_R}$ against heavy Majorana neutrino mass $M_1$ in TeV For NH and IH.} \label{fig1}
\end{figure}

Where,
 ${M_{RR}}^{(diag)}=diag(M_1, M_2, M_3)$.
We have fine tuned the dimensionless parameter, $\gamma\sim 10^{-10}$. The variation of the RH gauge boson mass with heavy RH neutrino mass as shown in
fig (\ref{fig1}), corresponds to the condition $M_{W_R}>M_N $. As previously mentioned we have considered three different values of the $SU(2)_R$ 
breaking scale, $v_R$ for our further analysis, specifically 5 TeV, 10 TeV and 18 TeV respectively, which will be useful to study the common parameter space
of the phenomenon we have considered, i.e., BAU, NDBD, LFV. The left handed gauge boson is $M_{W_L}= 80$ GeV and determined the RHS of equation terms of lightest neutrino mass by varying the Majorana phases from 0 to 2$\pi$. By considering a very tiny mass splitting of the Majorana masses
$M_1$ and $M_2$ as per requirement of resonant leptogenesis, we equated both sides of equation (\ref{eqa5}) and obtained  $M_1$, $M_2$ and  $M_3$,
where, $ M_1\approx M_2$.
\par We considered the lepton number violating and CP violating decays of two heavy RH Majorana neutrinos, $N_1$ and  $N_2$ via the decay modes,  $N_i\rightarrow l+\phi^{c} $ and its 
CP conjugate process,  $N_i\rightarrow l^c+\phi $, $i=1, 2$. Firstly, we determined the leptonic CP asymmetry, $\epsilon_1$ and $\epsilon_2$ using equation (\ref{eqa9})
where $	Y_\nu=\frac{M_D}{v}$, v being the VEV of Higgs bidoublet and is 174 GeV. The decay rates in equation (\ref{eqa9}) can be obtained using equation (\ref{eqa10}). 
The Dirac mass, $m_D$ as mentioned before is not of any specific texture, but we have obtained 
it from the type I seesaw equation in which we have considered the light neutrino mass $M_{LL}$ and the heavy right handed Majorana neutrino mass to be known, which
satisfies the current neutrino oscillation data.

\par The CP violating asymmetries $\epsilon_1$ and $\epsilon_2$ can give rise to a net lepton number asymmetry, provided the expansion rate of the universe is larger than
$\Gamma_1$ and $\Gamma_2$. The net baryon asymmetry is then calculated using \cite{zing}\cite{BA},

\begin{equation}\label{eqa10}
 \eta_B \approx -0.96 \times 10^{-2} \sum_i \left(k_i\epsilon_i\right),
\end{equation}

\begin{figure}[h!]
\includegraphics[width=0.46\textwidth]{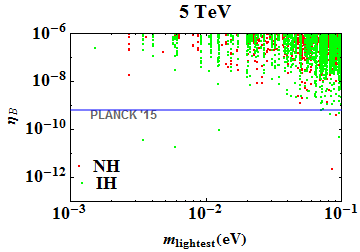}
\includegraphics[width=0.46\textwidth]{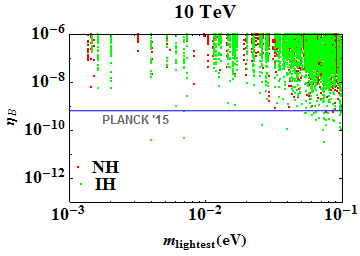}
\includegraphics[width=0.46\textwidth]{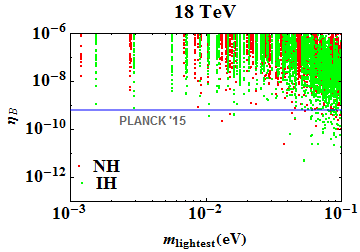}
\caption{BAU as a function of lightest neutrino mass, $m_1/m_3$ (in eV)for NH/IH. The blue solid line represents the observed BAU in PLANCK '15\cite{planck} for different values
of RH gauge boson mass, 5, 10 and 18 TeV respectively. } \label{fig2}
\end{figure}

$k_1$ and $k_2$ being the efficiency factors measuring the washout  effects linked with the out of equillibrium decay of $N_1$ and $N_2$. We can define the parameters,
$K_i\equiv \frac{\Gamma_i}{H}$ at temperature, $T= M_i$, $H\equiv \frac{1.66 \sqrt{g_*}T^2}{M_Planck}$ is the  Hubble's constant with $g_*\simeq$ 107 and 
$M_{Planck}\equiv 1.2 \times 10^{19} GeV$ is the Planck mass. The decay width can be estimated using equation (\ref {eqa8}). 
For simplicity, the efficiency factors, $k_i$ can be calculated using the formula \cite{K},
\begin{equation}\label{eqa11}
 k_1\equiv k_2\equiv \frac{1}{2}{\left(\sum_i K_i\right)}^{-1.2},
\end{equation}

 which holds validity for two nearly degenerate heavy Majorana  masses and $ 5\leq K_i\leq 100 $. We have used the formula \ref{eqa10} in calculating the baryon asymmetry.
 The result is shown as a function of lightest neutrino mass by varying the Majorana phases from 0 to 2 $\pi$ in fig (\ref{fig2}) for different values of RH gauge boson mass.
It is evident from the figure that the cosmologicl observed BAU from RL can be obtained for varying gauge boson mass $M_{W_R}$, distinctively, 5, 10 and 18 TeV in our
case, which is in accordance to several prior works. In the case of mass hierarchy, IH seems to give better results in our analysis. The required amount of BAU 
is perceived for lightest neutrino mass of around (0.05-0.1) eV. For $M_{W_R}=$ 18 TeV, greater parameter space satisfies the observed BAU than for 5 TeV.

\subsection{NDBD from heavy RH neutrino not so large left-right mixing}

In LRSM, owing to the presence of several new heavy particles, many new contributions arises to NDBD amplitudes. In a previous work (second reference of\cite{X})
we have considered the new physics contributions coming from the ones mediated by ${W_R}^-$ and $\Delta_R$ respectively. In the present work,
besides the heavy RH neutrino contribution coming from the
exchange of $W_R$ bosons, we also considered the momentum dependent mechanisms also, i.e., the $\lambda$ and $\eta$ contributions to NDBD due to gauge boson mixing since
we have seen non negligible contributions from these momentum dependent mechanisms in our case. 

The effective neutrino mass corresponding to the heavy RH neutrino contribution from the exchange of $W_R$ gauge bosons is given by,

\begin{equation}
 \rm  M^N_{eff}=p^2\frac{{M_{W_L}}^4}{{M_{W_R}}^4}\frac{{U_{Rei}}^*2}{M_i} .
\end{equation}

\par Here, $ \rm <p^2> = m_e m_p \frac{M_N}{M_\nu}$ is the typical momentum exchange of the process, where $ \rm m_p$ and $ \rm m_e$ are the mass of the proton and electron respectively 
and $ \rm M_N$ is the NME corresponding to the RH neutrino exchange. The allowed value of p (the virtuality of the exchanged neutrino) is in the range
$\sim $ (100-200) MeV. In our analysis, we have taken p$\simeq$180 MeV \cite{ndbddb2}. As in case of BAU, herein, we have considered different values of
$M_{W_R}$, namely, 5, 10 and 18 TeV respectively. $U_{Rei}$ are the first row elements of the diagonalizing matrix of the heavy right handed Majorana mass matrix 
$ \rm M_{RR}$ and  $ \rm M_i$ is its mass eigenvalues, $ \rm M_i$.

 \begin{itemize}
  \item In case of $\lambda$ contribution, the particle physics parameter that measures the lepton number violation is given by,
  \begin{equation}
   \left|\eta_\lambda\right|=\left(\frac{M_{W_L}}{M_{W_R}}\right)\left|\sum_i U_{e_i} T^*_{e_i}\right|.
  \end{equation}
\item While the $\eta$ contribution to NDBD due to $W_L-W_R$ mixing is described by the parameter, $\tan \zeta$, as in equation (\ref{eqx24}),
with particle physics parameter,
  \begin{equation}
    \left|\eta_\eta\right|= \tan \zeta \left|\sum_i U_{ei} T^*_{ei}\right|.
  \end{equation}
   \end{itemize}
 In the above equations, $U_{e_i}$ represents the elements of  the matrix  as defined by equation (\ref{eqx19}), and T is represented by equation (\ref{eqx22}), the term 
 $\left|\sum_i U_{ei} T^*_{ei}\right|$ can be simplified to the form $ -\left[M_DM_{RR}^{-1}\right]_{ee}$ (as in second reference of \cite{lfv1}).
 $V_\nu$ in the expression for T is the diagonalizing matrix of $M_\nu$. The effective Majorana neutrino mass due to $\lambda$  and $\eta$ contribution is thus given by,
\begin{equation}
 M^\lambda_{eff}=\frac{\eta_\lambda}{m_e} ,  \\ M^\eta_{eff}=\frac{\eta_\eta}{m_e}.
\end{equation}
 
The half lives corresponding to these effective mass values is given by,

\begin{equation}
 \left[{T_{\frac{1}{2}}}^{0\nu}\right]^{-1}=G^{0\nu}(Q,Z){\left|M^{0\nu}_N\right|}^2\frac{{\left|M^N_{eff}\right|_N}^2}{{m_e}^2},
\end{equation}
\begin{equation}
 \left[{T_{\frac{1}{2}}}^{0\nu}\right]^{-1}=G^{0\nu}(Q,Z){\left|M^{0\nu}_\lambda\right|}^2\frac{{\left| M^\lambda_{eff}\right|_N}^2}{{m_e}^2},
\end{equation}
\begin{equation}
 \left[{T_{\frac{1}{2}}}^{0\nu}\right]^{-1}=G^{0\nu}(Q,Z){\left|M^{0\nu}_\eta\right|}^2\frac{{\left| M^\eta_{eff}\right|_N}^2}{{m_e}^2},
\end{equation}

where, $G^{0\nu}$ and $\left|M^{0\nu}\right|$ represents the phase space factor and the nuclear matrix elements of the processes which holds different values as in \cite{nme}

\begin{figure}[h!]
\includegraphics[width=0.46\textwidth]{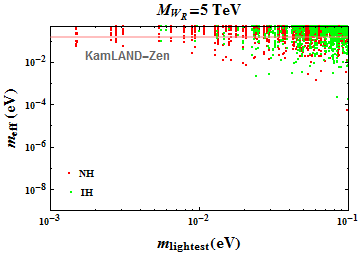}
\includegraphics[width=0.46\textwidth]{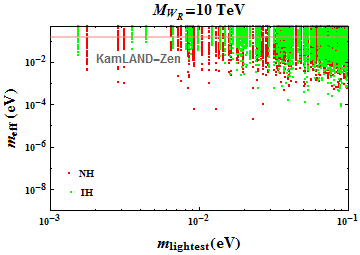}
\includegraphics[width=0.46\textwidth]{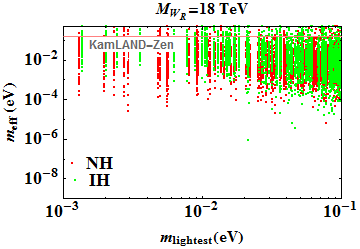}
\caption{Effective Majorana mass for 0$\nu\beta\beta$ as a function of lightest neutrino mass, for new physics contribution coming from RH $\nu$ for both NH and IH. The 
pink solid line represents the KamLAND-Zen upper bound on the effective neutrino mass.} \label{fig3}
\end{figure}

\begin{figure}[h!]
\includegraphics[width=0.44\textwidth]{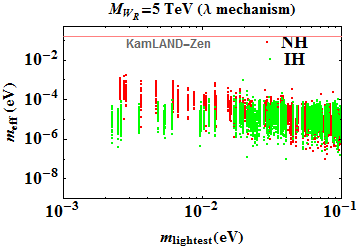}
\includegraphics[width=0.44\textwidth]{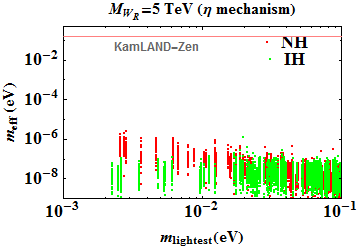}
\includegraphics[width=0.44\textwidth]{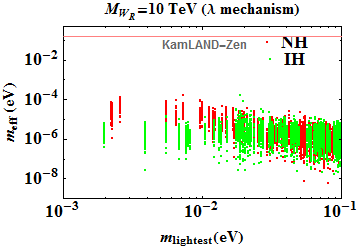}
\includegraphics[width=0.44\textwidth]{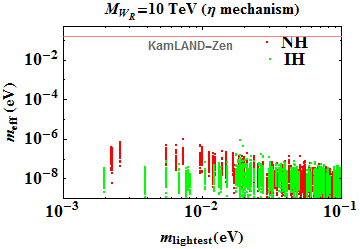}
\includegraphics[width=0.44\textwidth]{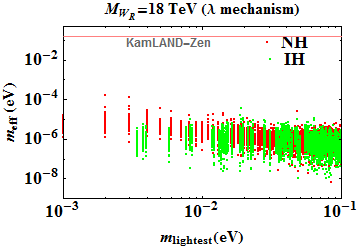}
\includegraphics[width=0.44\textwidth]{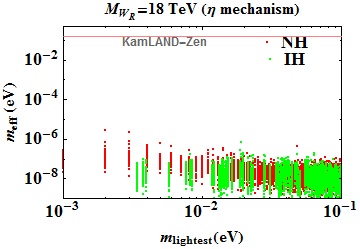}
\caption{Effective Majorana mass for 0$\nu\beta\beta$ as a function of lightest neutrino mass, for new physics contribution coming from $\lambda$ (left figures)and
$\eta$ mechanisms( right figures) for NH and IH for different RH gauge boson masses. The 
pink solid line represents the KamLAND-Zen upper bound on the effective neutrino mass.} \label{fig4}
\end{figure}

\begin{figure}[h!]
\includegraphics[width=0.43\textwidth]{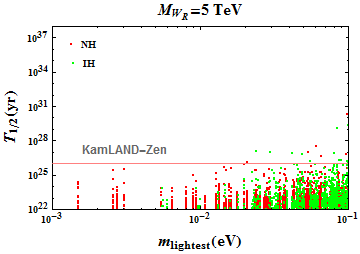}
\includegraphics[width=0.43\textwidth]{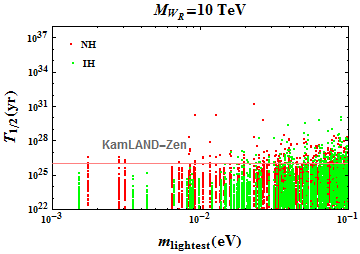}
\includegraphics[width=0.43\textwidth]{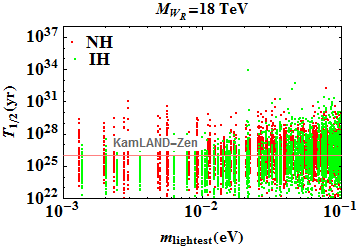}
\caption{Half life for 0$\nu\beta\beta$ as a function of lightest neutrino mass for NH and IH
for heavy RH neutrino contribution. The horizontal pink line  represents the KamLAND-Zen  lower bound on the half life of NDBD.} \label{fig5}
\end{figure}

\begin{figure}[h!]
\includegraphics[width=0.43\textwidth]{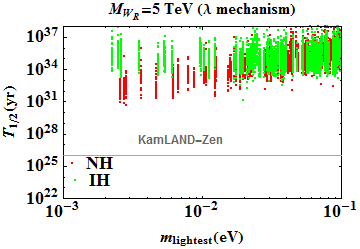}
\includegraphics[width=0.43\textwidth]{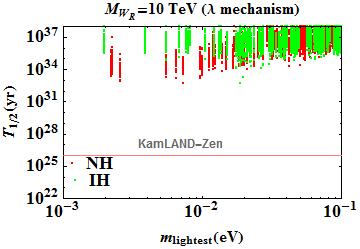}
\includegraphics[width=0.43\textwidth]{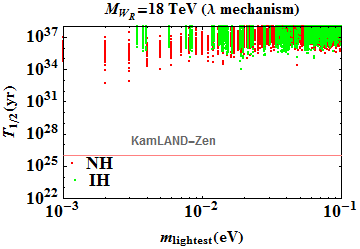}
\caption{Half life for 0$\nu\beta\beta$ as a function of lightest neutrino mass for NH and IH
for $\lambda$ mechanism. The horizontal line represents the KamLAND-Zen lower bound on the half life of NDBD.} \label{fig6}
\end{figure}

\begin{figure}[h!]
\includegraphics[width=0.40\textwidth]{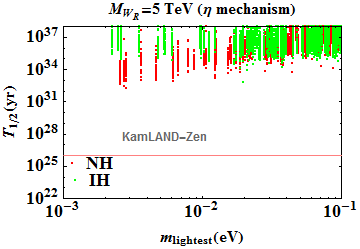}
\includegraphics[width=0.40\textwidth]{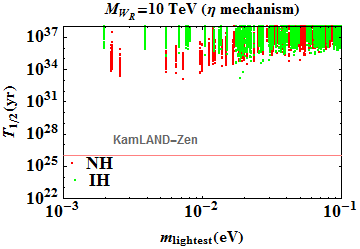}
\includegraphics[width=0.40\textwidth]{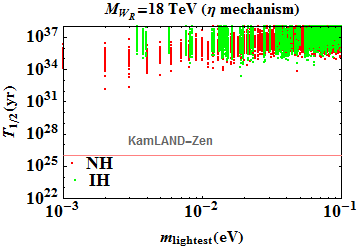}
\caption{Half life for 0$\nu\beta\beta$ as a function of lightest neutrino mass for NH and IH
for $\eta$ mechanism. The horizontal line represents the KamLAND-Zen lower bound on the half life of NDBD.} \label{fig7}
\end{figure}

 Fig (\ref{fig3}) to (\ref{fig7}) shows the effective mass and half life governing NDBD from RH neutrino, $\lambda$ and $\eta$ contribution against the lightest neutrino mass. 
 For new physics contribution coming from purely RH current, the effective mass governing NDBD is consistent with the experimental results as propounded by
 KamLAND-Zen for all the cases ($M_{W_R}$= 5, 10, 18 TeV) although better results is obtained for 18 TeV. It is not much dependent on the mass hierarchy.
 Whereas, for NDBD contributions from $\lambda$ and $\eta$ mechanisms, the effective mass is found to be within the experimental limit but of lower magnitude 
 than the RH neutrino contributions. We have seen that $\eta$ contribution ($10^{-6}-10^{-8}$)eV is around two orders of magnitude less than the $\lambda$
 contribution ($10^{-4}-10^{-6}$)eV in all the cases irrespective of the mass hierarchies. Similar results are obtained for the half lives of the process.

 \par
\begin{figure}[h!]
\includegraphics[width=0.37\textwidth]{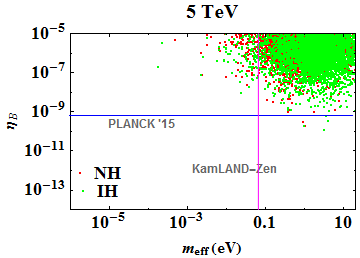}
\includegraphics[width=0.37\textwidth]{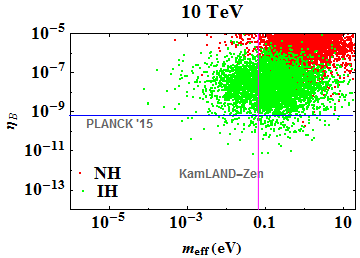}
\includegraphics[width=0.37\textwidth]{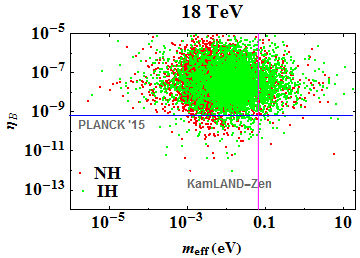}
\caption{BAU against effective Majorana neutrino mass for RH $\nu$ contribution.The solid blue and pink line represents the observed BAU and the KAMLAND upper 
bound on effective Majorana neutrino mass.} \label{fig8}
\end{figure}

\begin{figure}[h!]
\includegraphics[width=0.37\textwidth]{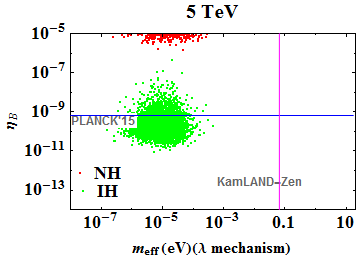}
\includegraphics[width=0.37\textwidth]{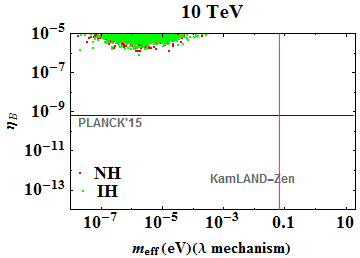}
\includegraphics[width=0.37\textwidth]{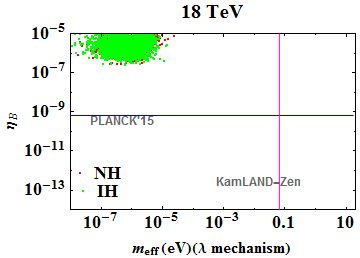}
\caption{BAU against effective Majorana neutrino mass (for $\lambda$ mechanism) . The solid blue and pink line represents the observed BAU and the KAMLAND upper 
bound on effective Majorana neutrino mass.} \label{fig9}
\end{figure}

\begin{figure}[h!]
\includegraphics[width=0.37\textwidth]{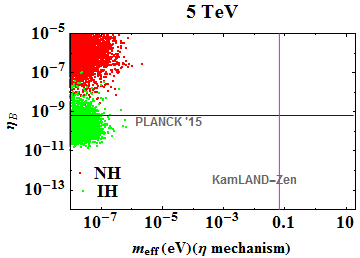}
\includegraphics[width=0.37\textwidth]{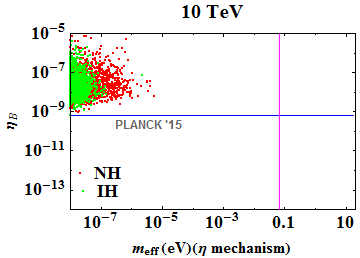}
\includegraphics[width=0.37\textwidth]{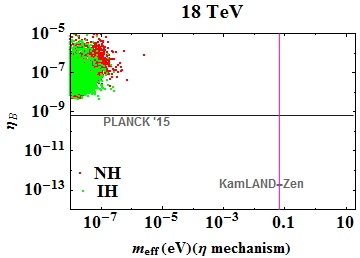}
\caption{BAU against effective Majorana neutrino mass  (for $\eta$ mechanism).The solid blue and pink line represents the observed BAU and the KAMLAND upper 
bound on effective Majorana neutrino mass.} \label{fig10}
\end{figure}

 Fig (\ref{fig8}) to (\ref{fig10}) shows the correlation of NDBD and BAU for the different contributions. It is seen that BAU and NDBD (for RH $\nu$
 contribution) can simultaneously satisfy the experimental results for $M_{W_R}=$ 10 and 18 TeV in our case, although for 10 TeV case only IH is consistent 
 with the experimental bounds. As far as the mixed contributions are concerned, a common parameter space for NDBD and BAU is observed only for RH gauge boson mass
 to be 5 TeV and for IH only.

\subsection{Lepton Flavor Violation}

In our analysis, we further studied the LFV processes, $\mu\rightarrow 3e $  and  $\mu\rightarrow e\gamma $ and correlated the branching ratios(BR) with the lightest 
neutrino mass and the atmospheric mixing angle respectively as in our previous work ( second reference of \cite{X}).
For calculating the BR, we used the expression given in equation (\ref{eq43}) . The lepton Higgs coupling $h_{ij}$ in (\ref{eq44}) can be computed explicitly for a given RH neutrino mass matrix as shown in
equation (\ref{eqa5}) by diagonalizing the RH neutrino mass matrix and obtaining the mixing matrix element, $V_i$ and the eigenvalues $M_i$. For evaluating $M_{RR}$, we 
need to know ${ M_\nu}^{II} $, as evident from equation (\ref{eqa5}). We computed ${ M_\nu}^{II} $ from equation (\ref{eqa6}). For determining the BR for $\mu\rightarrow 3e$, 
we imposed the best fit values of
the parameters, $\rm{\Delta m_{sol}}^2$, $\rm{\Delta m_{atm}}^2$, $\rm\delta$, $\rm\theta_{13}$, $\rm\theta_{23}$, $\rm\theta_{12}$ in $\rm M_\nu$ .
The numerical values of $\rm{ M_\nu}^{I} $ can be computed considering TBM mixing pattern in our case. Thus, we get $\rm{ M_\nu}^{II} $ as a function of the 
parameters $\rm\alpha,\beta$ and $\rm m_{lightest}$. 
Then varying both the Majorana phases, $\rm\alpha, \beta$ from 0 to 2$\pi$, we obtained $\rm{ M_\nu}^{II} $ as a function of $m_{lightest}$. 
Similarly, for $\mu\rightarrow e\gamma$
 we substituted the values of the lightest mass (m1/m3)for(NH/IH) as (0.07eV/0.065eV) and best fit values for the parameters ${\Delta m_{sol}}^2$, ${\Delta m_{atm}}^2$,
 $\rm\delta$, $\rm\theta_{13}$, while varying both the Majorana phases, $\alpha, \beta$ from 0 to 2$\pi$ and thus 
obtained $\rm{ M_\nu}^{II} $ and hence $\rm M_{RR}$ as a function of the atmospheric mixing angle $\theta_{23}$. Thus BR can be obtained as a function of $\sin^2\theta_{23}$
from equation (\ref{eq43}). We have varied the value of $\rm\sin^2\theta_{23}$ in its 3\rm$\sigma$ range as in \cite{bestfit} and the lightest neutrino mass
from $10^{-3}$ to $10^{-1}$ and obtained the values of BR. Like the previous cases  (BAU and NDBD), we have considered three values of the 
right handed gauge boson mass, 5 TeV, 10 TeV and 18 TeV respectively and different results have been obtained for these different values. 
\par
\begin{figure}[h!]
\includegraphics[width=0.38\textwidth]{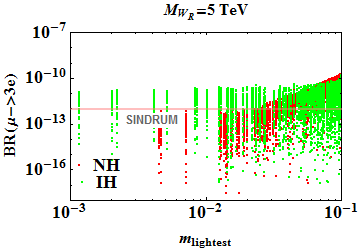}
\includegraphics[width=0.38\textwidth]{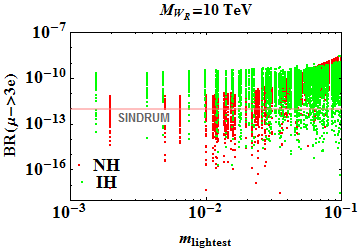}
\includegraphics[width=0.38\textwidth]{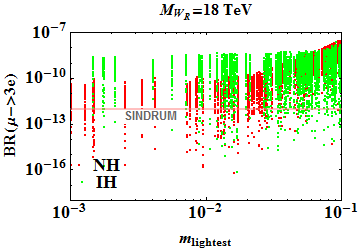}
\caption{ BR for $\rm \mu\rightarrow 3e $ shown as a function of the lightest neutrino mass. The solid pink line represents the limit of BR as given
by SINDRUM experiment.} \label{fig11}
\end{figure}

\begin{figure}[h!]
\includegraphics[width=0.36\textwidth]{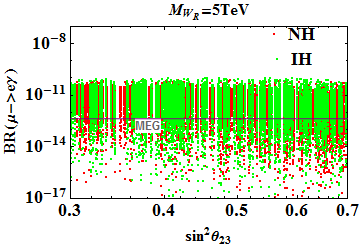}
\includegraphics[width=0.37\textwidth]{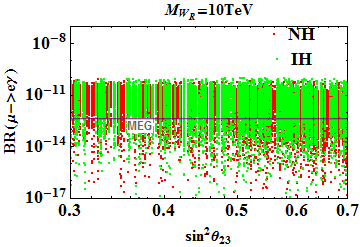}
\includegraphics[width=0.41\textwidth]{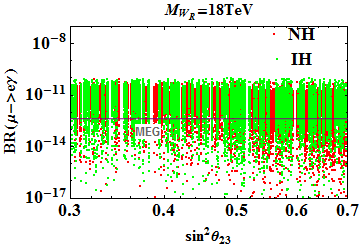}
\caption{BR for $\rm \mu\rightarrow e\gamma $ shown as a function of the atmospheric mixing angle. The horizontal solid line shows the limit of BR as
given by MEG experiment.} \label{fig12}
\end{figure}

 The variation is shown in figure (\ref{fig11}) and (\ref{fig12}) for both NH and IH. It is obvious from the figures that for both the LFV process, a good amount
of parameter space is consistent with the experimental results for the different RH gauge boson mass we have considered i.e. 5, 10 and 18 TeV.

We have shown a summarized form of our results in tabular form in table \ref{t3}.

\clearpage


\clearpage



 \begin{table}[h]
\centering
\begin{tabular}{||c| c| c| c||}
\hline
OBSERVABLES& 5 TeV NH (IH) & 10 TeV NH (IH)  & 18 TeV NH (IH)\\ \hline
NDBD($N_R$)  & $ \checkmark (\checkmark)$ & $ \checkmark (\checkmark)$&$ \checkmark (\checkmark)$ \\ \hline
NDBD$(\lambda)$ &$\checkmark(\checkmark)$ & $\checkmark(\checkmark)$&$\checkmark (\checkmark)$ \\ \hline
NDBD$(\eta)$ &$\checkmark(\checkmark)$ & $\checkmark (\checkmark)$&$\checkmark (\checkmark)$ \\ \hline
BAU & $ \checkmark (\checkmark)$ & $ \checkmark (\checkmark)$&$ \checkmark (\checkmark)$ \\ \hline
BAU and NDBD($N_R$) & $ \times (\times)$ & $ \times (\checkmark)$&$ \checkmark (\checkmark)$ \\ \hline
BAU and NDBD$(\lambda)$ & $ \times (\checkmark)$ & $ \times (\times)$&$ \times (\times)$ \\ \hline
BAU and NDBD$(\eta)$ &$ \times (\checkmark)$ & $ \times (\times)$&$ \times (\times)$ \\ \hline
BR$(\mu\rightarrow 3e )$& $ \checkmark (\checkmark)$ & $ \checkmark (\checkmark)$&$ \checkmark (\checkmark)$ \\ \hline
BR$(\mu\rightarrow e\gamma)$ &$ \checkmark (\checkmark)$ & $ \checkmark(\checkmark)$&$ \checkmark (\checkmark)$ \\ \hline
\end{tabular}
\caption{summarized form of the results for NDBD, BAU, LFV for both NH and IH. The $\checkmark$ and $\times$ symbol are used to denote if the observables are (not are) in
the current experimental limit}\label{t3}
\end{table}

\clearpage

\section{DISCUSSION AND CONCLUSION}{\label{sec:level6}}
While calculating the NDBD contribution and BAU we concentrated on an important issue that whether both the phenomena 
can be correlated in TeV scale or not. As addressed  by the author in \cite{18tev} TeV-scale LRSM, there are complications due to the presence of RH gauge 
interactions that contribute to the dilution and washout of the primordial
lepton asymmetry generated via resonant leptogenesis. Combined with the dilution effects from inverse decays and entropy, this
implies that even for maximal CP asymmetry the observed baryon-to-photon ratio can be obtained only if $M_{W_R} \geq18$ TeV. 
They have 
basically focussed on the possibilities of falsification of leptogenesis owing to the possible experimental observation of RH gauge boson mass of around
$(3-5)$ TeV.
But in the recent papers \cite{3tev} \cite{10tev} authors have taken up this issue and claim that one can generate the baryon asymmetry within the experimental
limit even if  RH gauge boson mass is as low as 5 TeV. In their work, instead of assuming maximal CP asymmetry, they calculated the premordial
CP asymmetry as demanded by their specific neutrino fix. Furthermore, they have also shown the consistency of their model with other low energy constraints
like NDBD, LFV etc. thereby specifying the fact that just the possible observation of $W_R$ at LHC alone cannot falsify leptogenesis as a mechanism to generate 
matter- antimatter asymmetry of the universe. Since the main purpose of our work is to see if there is a common parameter space where we can establish a 
linkage between baryogenesis and the low scale phenomenon like NDBD and LFV, we have done a phenomenological study of these phenomenon at a TeV
scale LRSM considering some specific values of RH  Gauge boson mass, 5 TeV, 10 TeV and 18 TeV (as found separately in the earlier works) and check the 
consistency of the previous results. 
Based on our study, we could arrive at the following conclusions,

\begin{itemize}
 \item For a low scale model independent seesaw model, one can account for successful leptogenesis and also the constraints that comes after regarding mass of the 
 RH gauge bosons is that larger parameter space for BAU with the observed cosmological value is obtained for $M_{W_R}=$ 18 TeV than for 5 TeV. 
 
 \item New Physics contributions to NDBD in TeV scale LRSM for different $M_{W_R}$ shows that dominant contribution comes from the exchange of RH gauge boson rather
 than the mixed, LH-RH gauge boson mixing contributions. The $\lambda$ contributions to NDBD is a bit suppressed owing to the less Yukawa coupling and not so 
 large left-right mixing in our analysis while $\eta$ contribution is further supressed by two orders of magnitude that the $\lambda$ contribution.
 
 \item It is possible to obtain a common parameter space for both NDBD and BAU. This corresponds  to the NDBD contribution coming from the heavy RH neutrino for both NH
 and IH. However, in this case better results are obtained for 18 TeV RH gauge boson mass. Whereas, as far as the
 the momentum dependent $\lambda$ and $\eta$ mechanisms are concerned, both NDBD and BAU can be simultaneously explained  for $M_{W_R}=$5 TeV or $\leqslant$ 10 TeV and only
 for IH.
 
 \item Sizeable implications for other low energy observable,charged  LFV of the processes, $\rm \mu\rightarrow 3e $ and $\rm \mu\rightarrow e\gamma$ are obtained for a
 minimal TeV scale LRSM which simultaneously accounts for BAU and NDBD.
 \par For LFV, the BR prediction for $\rm \mu\rightarrow e\gamma$ is not much dependent on the atmospheric mixing angle, $\theta_{23}$. 
\end{itemize}

Having done an extensive study of several of the earlier works, we have found that our results are in accordance with the previous works where low scale phenomena are discussed.
That successful leptogenesis can be 
found within the vicinity of the experimental limit for RH gauge boson mass as low as 5 TeV and is not much dependent on the mass hierachy, NH or IH. However, both low scale BAU and
effective mass governing NDBD can be simultaneously obtained for only some
parameter space that depends on the mass hierarchy and the $W_R$ mass as mentioned in the above points. Notwithstanding a more detailed study is preferred in order to give 
a strong concluding remark.
\clearpage
\section{APPENDIX}{\label{sec:level7}}
\textbf{Determination of $\rm M_D$:}\\
From type I SS term,  $ { M_\nu}^{I}\approx -M_DM^{-1}_{RR}M^T_D$

Again, ${ M_\nu}^{I}= U_{(TBM)}U_{Maj}X{M_\nu}^{(diag)}{U_{Maj}}^T{U_{(TBM)}}^T$

\begin{equation}\label{b2}
 M_{RR}=\frac{1}{\gamma}{\left(\frac{v_R}{M_{W_L}}\right)}^2{ M_\nu}^{II}
\end{equation}

Where, ${ M_\nu}^{II}=U_{PMNS}{M_\nu}^{(diag)} {U_{PMNS}}^T- U_{(TBM)}U_{Maj}X{M_\nu}^{(diag)}{U_{Maj}}^T{U_{(TBM)}}^T$.
Considering, X=0.5, $M_{W_L}= 80$ GeV, $v_R= 5$ TeV (for one case only) ,and expressing ${M_\nu}^{(diag)}$  in terms of lightest
neutrino mass, $m_1(m_3)$ for NH (IH), we obtained $M_{RR}$ varying the Majorana phases $\alpha$ and $\beta$ from 0 to 2$\pi$ and lightest neutrino mass
from $10^{-3}$ to $10^{-1}$.

We have considered $M_D$ as,
\begin{equation}\label{b3}
\rm M_{D}=\left[\begin{array}{ccc}
a_1&a_2&a_3\\
a_2&a_4&a_5\\
a_3&a_5&a_6
\end{array}\right],
\end{equation}

which is $\mu-\tau$ symmetric. Equating both sides of type I seesaw equation and solving for $a_1,a_2,a_3,a_4,a_5,a_6$, we obtain 
the matrix elements of one of the $M_D$ of the form,

\begin{equation}\label{b4}
\rm M_{D}=\left[\begin{array}{ccc}
24776.2+122368.i&70524.8+76561.i&-12687.1+21472.4i\\
70524.8+76561.i&14308.4+138730.i&-45802.3-46293.4i\\
-12687.1+21472.4i&-45802.3-46293.4i&87313.6+158166.i
\end{array}\right],
\end{equation}
which we have implemented for our further analysis.

\textbf{Elements of the type II Seesaw mass matrix:}

\begin{equation}
 S_{11}=\left(c^2_{12}c^2_{13}-X{c_{12}^2}^{TBM}\right)m_1+e^{2i(\beta-\delta)}s^2_{13}m_3+\left(c^2_{13}s^2_{12}-X{s_{12}^2}^{TBM}\right)e^{2i\alpha}m_2
\end{equation}
\begin{equation}
\begin{split}
S_{12}=\left(-c_{12}c_{13}c_{23}s_{12}-c^2_{12}c_{13}s_{13}s_{23}e^{i\delta}+X{c_{12}^{TBM}}{c_{23}^{TBM}}{s_{12}^{TBM}}\right)m_1+\\
\left(-c_{13}s_{12}c_{12}c_{23}e^{2i\alpha}-c_{13}s^2_{12}s_{13}s_{23}e^{i(2\alpha+\delta)}+X{c_{12}^{TBM}}{c_{23}^{TBM}}{s_{12}^{TBM}}e^{2i\alpha}\right)m_2+\\
\left(c_{13}s_{13}s_{23}e^{i(2\beta-\delta)}\right)m_3
\end{split}
\end{equation}
\begin{equation}
\begin{split}
 S_{13}=\left(c^2_{12}c_{13}c_{23}s_{13}e^{i\delta}+s_{12}s_{23}c_{12}c_{13}-X{c_{12}^{TBM}}{s_{12}^{TBM}}{s_{23}^{TBM}}\right)m_1+\\
 \left(-c_{13}s_{12}c_{23}s_{12}s_{13}e^{i(2\alpha+\delta)}-X{c_{12}^{TBM}}{s_{12}^{TBM}}{s_{23}^{TBM}}e^{2i\alpha}\right)m_2+\\
 \left(e^{i(2\beta-\delta)}c_{13}c_{23}s_{13}\right)m_3
 \end{split}
\end{equation}
\begin{equation}
\begin{split}
 S_{21}=\left(-c_{12}c_{13}c_{23}s_{12}-c^2_{12}c_{13}s_{13}s_{23}e^{i\delta}+X{c_{12}^{TBM}}{c_{23}^{TBM}}{s_{12}^{TBM}}\right)m_1+\\
 \left(c_{13}s_{12}c_{12}c_{23}e^{2i\alpha}-s^2_{12}s_{13}s_{23}c_{13}e^{i(2\alpha+\delta)}+X{c_{12}^{TBM}}{c_{23}^{TBM}}{s_{12}^{TBM}}e^{2i\alpha}\right)m_2\\
 \left(e^{i(2\beta-\delta)}c_{13}s_{23}s_{13}\right)m_3
 \end{split}
\end{equation}
\begin{equation}
 \begin{split}
  S_{22}=\left({\left(c_{23}s_{12}-e^{i\delta}c_{12}s_{13}s_{23}\right)}^2-X{c_{23}^2}^{TBM}{s_{12}^2}^{TBM}\right)m_1+\\
  \left(-X{c_{12}^2}^{TBM}{c_{23}^2}^{TBM}+{\left(-c_{12}c_{23}-e^{i\delta}s_{12}s_{13}s_{23}\right)}^2\right)m_2e^{2i\alpha}+\\
  \left(c^2_{13}s^2_{23}-X{s_{23}^2}^{TBM}e^{2i\beta}\right)m_3
  \end{split}
\end{equation}
\begin{equation}
 \begin{split}
S_{23}= \left(\left(-c_{12}c_{23}s_{13}e^{i\delta}+s_{12}s_{23}\right)\left(-c_{23}s_{12}-e^{i\delta}c_{12}s_{13}s_{23}\right)+X{c_{23}^{TBM}}{s_{12}^2}^{TBM}{s_{23}^2}^{TBMu}\right)m_1+\\
 \left(\left(-e^{i\delta}c_{23}s_{12}s_{13}+c_{12}s_{23}\right)\left(-c_{12}c_{23}-e^{i\delta}s_{12}s_{13}s_{23}\right)+X{c_{12}^2}^{TBM}{c_{23}^{TBM}}{s_{23}^{TBM}}\right)m_2e^{2i\alpha}+\\
 \left(c^2_{13}c_{23}s_{23}e^{2i\beta}-{c_{23}^{TBM}}{s_{23}^{TBM}}\right)m_3
 \end{split}
\end{equation}
\begin{equation}
 \begin{split}
S_{31}=\left(c^2_{12}c_{13}c_{23}s_{13}e^{i\delta}+s_{12}s_{23}c_{12}c_{13}-X{c_{12}^{TBM}}{s_{12}^{TBM}}{s_{23}^{TBM}}\right)m_1+\\
  \left(c_{13}s^2_{12}e^{i\delta}c_{23}s_{13}+c_{12}s_{23}c_{13}s_{12}e^{2i\alpha}-X{c_{12}^{TBM}}{s_{12}^{TBM}}{s_{23}^{TBM}}\right)m_2e^{2i\alpha}+\\
  \left(e^{2i\beta-i\delta}c_{13}c_{23}s_{13}\right)m_3
 \end{split}
\end{equation}
\begin{equation}
 \begin{split}
S_{32}=\left(\left(-e^{i\delta}c_{12}c_{23}s_{13}+s_{12}s_{23}\right)\left(-c_{23}s_{12}-e^{i\delta}c_{12}s_{13}s_{23}\right)+{c_{23}^{TBM}}{s_{12}^2}^{TBM}{s_{23}^{TBM}}\right)m_1\\
 \left(\left(-e^{i\delta}c_{23}s_{12}s_{13}+c_{12}s_{23}\right)\left(-c_{12}c_{23}-e^{i\delta}s_{12}s_{13}s_{23}\right)+X{c_{12}^2}^{TBM}{c_{23}^{TBM}}{s_{23}^{TBM}}\right)e^{2i\alpha}m_2\\
  \left(c^2_{13}c_{23}s_{23}-X{c^{TBM}}_{23}{s^{TBM}}_{23}\right)e^{2i\beta}m_3
 \end{split} 
\end{equation}
\begin{equation}
 \begin{split}
S_{33}=\left({\left(-e^{i\delta}c_{12}c_{23}s_{13}+s_{12}s_{23}\right)}^2-X{s_{12}^2}^{TBM}{s_{23}^2}^{TBM}\right)m_1+\\
  \left({\left(-e^{i\delta}c_{23}s_{12}s_{13}+c_{12}s_{23}\right)}^2-X{c_{12}^2}^{TBM}{s_{23}^2}^{TBM}\right)e^{2i\alpha}m_2+\\
  \left(c^2_{13}c^2_{23}-{c_{23}^2}^{TBM}\right)e^{2i\beta}m_3
 \end{split}
\end{equation}
Where, $\rm {c_{ij}^{TBM}}= \cos\theta_{ij}^{TBM}$, $\rm {s_{ij}^{TBM}}=\sin\theta_{ij}^{TBM}$ represents the mixing angles for TBM  neutrino mass matrix.

\end{document}